\documentclass[prc,aps,final,twocolumn,nofootinbib]{revtex4-1}
\usepackage{graphicx}
\usepackage{mathrsfs}
\usepackage{bm}
\usepackage{colordvi}
\usepackage[normalem]{ulem}

\newcommand{\bel}[1]{\begin{equation}\label{#1}}

\usepackage{epsfig,psfrag}

\def\ramaSM{\vadjust{\vbox to 0pt{\vss \hbox to \hsize
{\hskip\hsize \quad $\Leftarrow$\quad {\it SM}\hss}\vskip3.5pt}}}
\def\ramaSL{\vadjust{\vbox to 0pt{\vss \hbox to \hsize
{\hskip\hsize \quad $\Leftarrow$\quad {\it SL}\hss}\vskip3.5pt}}}
\def\ramaSR{\vadjust{\vbox to 0pt{\vss \hbox to \hsize
{\hskip\hsize \quad $\Leftarrow$\quad {\it SR}\hss}\vskip3.5pt}}}
\def\rama{\vadjust{\vbox to 0pt{\vss \hbox to \hsize
{\hskip\hsize \quad $\Leftarrow$\quad
{$\Longleftarrow$}\hss}\vskip3.5pt}}}

\def\be{\begin{equation}}
\def\ee{\end{equation}}
\def\bea{\begin{eqnarray}}
\def\eea{\end{eqnarray}}

\def\hahat{\hat{H}}

\def\hahat0{\hat{H}_0}

\def\exp{\hbox{exp}}

\def\siml{\hbox{\kern.1em \lower.6ex \hbox{$\sim$} \kern-1.12em
          \raise.6ex \hbox{$<$} \kern.1em }}
\def\simg{\hbox{\kern.1em \lower.6ex \hbox{$\sim$} \kern-1.12em
          \raise.6ex \hbox{$>$} \kern.1em }}

\def\siml{\hbox{\kern.1em \lower.6ex \hbox{$\sim$} \kern-1.12em
 \raise.6ex \hbox{$<$} \kern.1em}}
\def\simg{\hbox{\kern.1em \lower.6ex \hbox{$\sim$} \kern-1.12em
 \raise.6ex \hbox{$>$} \kern.1em}}

\newcommand{\beqar}{\begin{eqnarray}}
\newcommand{\eeqar}[1]{\label{#1} \end{eqnarray}}

\begin{document}

\title{Simple approach to the chaos-order contributions
and symmetry breaking in
nuclear spectra}

\author{A.G. Magner}
\email{Email: magner@kinr.kiev.ua}
\author{A.I. Levon}
\email{Email: levon@kinr.kiev.ua}
\author{S.V. Radionov}
\email{Email: sergey.radionov18@gmail.com}
\affiliation{\it  Institute for Nuclear Research, 03680 Kyiv, Ukraine}
\bigskip
\date{October, 5th, 2018}
\bigskip

\begin{abstract}

The simple one-parameter nearest neighbor-spacing
distribution (NNSD) is suggested
for statistical analysis
of nuclear spectra.  This distribution is
derived within the Wigner-Dyson approach
in the linear approximation for the level repulsion density of
quantum states. The obtained NNSD gives  the
individual information on
the Wigner and Poisson contributions in agreement
with that of the statistical experimental distributions of
collective states in deformed nuclei.
Using this NNSD, one finds that the
symmetry breaking due to the fixing of projections
of the angular momentum of collective states enhances a chaos as a shift
of the NNSD from the Poisson to Wigner distribution behavior.

\end{abstract}

\maketitle

\noindent


\section{Introduction}
\label{introd}

Statistical analysis of  the quantum energy spectra for complex many-body systems
such as atomic nuclei is in fruitful progress
\cite{stokmann,haake,Zelev96,GuhrMulWeinden98,Weiden09,Mitch10,Gomez11}. 
Different statistical methods  have been
proposed to obtain information on the chaoticity versus regularity
in nuclear spectra
 \cite{Porter65,Pandey79,Brody81,Berry81,Bohigas84,Mehta04,Aberg02}.
To satisfy their statistical criteria,  the main idea
 was to compile  sequences of states
having the same quantum numbers, e.g.,
the angular momentum and parity in several nuclei
 \cite{Shrin90,Shrin91,Shrin04,Vidmar07}.
The short-range fluctuation properties in
experimental spectra
are analyzed usually in terms of the nearest-neighbor
spacing distributions (NNSDs).
For a quantitative measure of the
degree of chaoticity of the many-body dynamics,
the statistical probability density
$p(s)$ as a function
of  spacings $s$ between the nearest neighboring levels
can be derived within
the general Wigner-Dyson (WD) approach based
on the level repulsion density $g(s)$ (the units will be specified
below)
\cite{Gomez11,Porter65,Brody81,Mehta04,Aberg02},
\begin{equation}
p(s) = g(s)\;
\exp\left(-\int_0^s g(s')\;\mbox{d}s'\right)\;.
\label{WDgen}
\end{equation}
The order
is approximately associated with the Poisson 
dependence of $p(s)$ on
the spacing $s$ variable for
$g(s)$, that is independent of $s$.
The chaoticity can be related to the Wigner distribution,
as clearly follows
for $g(s)\propto s$ \cite{Wigner51}. An intermediate nature of spectra between these two limit statistics should be expected,
see for instance Refs.~
\cite{Pandey79,LenzHaake90,Rabson04,Gomez11,Schierenberg12}.

 The estimated values of parameters of 
 the NNSD shed light on the intermediate
 statistical situation with considered spectra.
The Brody NNSD \cite{Bro73} is
based on the expression for the level repulsion density
that interpolates
between the Poisson and the Wigner
distribution. Berry and Robnik  \cite{Berrob84}
derived the NNSD starting from the microscopic expression for the density
of levels of a system through its classical Hamiltonian.
Other one-parameteric distribution NNSDs
were suggested in Refs.~\cite{Izrail88,Izrail90,Gomez11,Jafar12}.

For
further studies
of the order-chaos properties of
nuclear systems, it might be worthwhile
to apply a simple  analytical approximation to
the level repulsion density $g(s)$ in Eq.~(\ref{WDgen}).
For analysis of the statistical properties
in terms of the mixed Poisson and Wigner distributions,
the linear WD (LWD) approximation to
the level repulsion density  $g(s)$
was suggested in Refs.
\cite{BMprc12,LMRprc17}.
It is the two-parameter LWD  (LWD2, see Appendix); in contrast,
e.g., to  the
one-parameter Brody approach \cite{Brody81}.
 The LWD (LWD2)
approximation, as based on a smooth analytical (linear)
function $g(s)$ of $s$,  can be justified
within the WD theory,  see also Ref.~\cite{Aberg02}.
Moreover, it gives more proper
information on the separate Poisson order-like and
Wigner chaos-like contributions.
The  LWD2 NNSD $p^{}_2(s)$ was applied
recently \cite{LMRprc17}
for a statistical study of experimental data
\cite{Lev94,Lev09,Lev13,Lev15,Spi13,Les02,Buc06,Mey06,Bet09,Spi17}
on the collective states in deformed nuclei
with a given angular momentum $I$
and parity $\pi$, and compared
with the Brody distribution \cite{Brody81}.
These results are in accordance with the works of  Shriner et al.
\cite{Shrin91,Shrin90}.
They are alternative to that for the nuclear states of a
single-particle nature; see, e.g.,
Ref.~\cite{Dietz17}.
To
derive the NNSD with one parameter from Eq.~(\ref{WDgen})
and, at the same time, keep
the same quantitative individual information of their order
and chaos contributions
is still  an open question.  In addition,
the attractive subject of the research is to learn statistical properties
of the new symmetry breaking phenomenon 
\cite{Mitch88,Paar90,Mitch04,Benzoni05,Agvaan09,Gomez11}, in particular,
  a vilation of the isospin symmetry and pair correlation breaking.
  Another attractive subject is
related to a symmetry breaking
due to a fixed
projection of the angular momentum $K$ in a nuclear collective motion
\cite{Paar90,Benzoni05,Gomez11}.

In the present work, we obtain the
probability distribution
$p^{}_1(s)$ with a single parameter
on the basis of the linear approximation (LWD1) to
a level repulsion density $g(s)$ in Eq.~(\ref{WDgen})
and compare with the previously presented
(Brody \cite{Bro73} and LWD2 \cite{LMRprc17}) approaches, and
with the new experimental data for
a symmetry breaking observation.
The statistical properties of the nuclear collective states obtained by
the NNSD $p^{}_1(s)$
are tested below using
the experimental results from
Refs.~
\cite{Lev94,Lev09,Lev13,Lev15,Spi13,Les02,Buc06,Mey06,Bet09,Spi17} 
for NNSDs fitted
by the LWD1. The LWD1 NNSD is applied
also for studying
the new symmetry breaking phenomenon \cite{Gomez11,Lev09,Lev13,Lev15}.

\section{Wigner-Dyson LWD approach}
\label{LWD1}

Key quantity in Eq.~(\ref{WDgen}) is
the  level repulsion  density  $g(s)$.
It is convenient
to consider $s$ in units of the average $D$ of distances
between levels, $s=S/D$, where $S$ is the
distance between neighbor levels
and $D$ is locally a mean distance between neighboring
levels in  usual energy units.

The experimental
data are always known within the finite spacing interval, and
both normalization conditions
(\ref{normcond1LWD2}) and (\ref{normcond2LWD2})
can be dependent on the upper integration limit $s_{\rm max}$.
Assuming, however, a good convergence over spacing variable $s$,
one can approximately simplify these conditions
for the probability distribution
$p(s)$ as function of the
dimensionless variable $s$ by expanding $s_{\rm max}$ to the infinity,
\begin{equation}
\int_0^{\infty} p(s)\;\mbox{d}s=1\;,
\label{normcond1}
\end{equation}
\begin{equation}
\int_0^{\infty}  s\; p(s)\;\mbox{d}s= 1\;.
\label{normcond2}
\end{equation}
For the Poisson and Wigner limits,
from Eq.\ (\ref{WDgen}) one has  the corresponding well known
distributions, which obey Eqs.~(\ref{normcond1}) and (\ref{normcond2}),
\begin{equation}
p^{}_{\rm P}(s)
=\exp\left(-s\right),~~ p^{}_{\rm W}(s)
=\left(\frac{\pi s}{2}\right)\;\exp\left(-\frac{\pi s^2}{4}\right).
\label{pois}
\end{equation}
%
\begin{figure*}
\includegraphics[width=0.75\textwidth,clip]{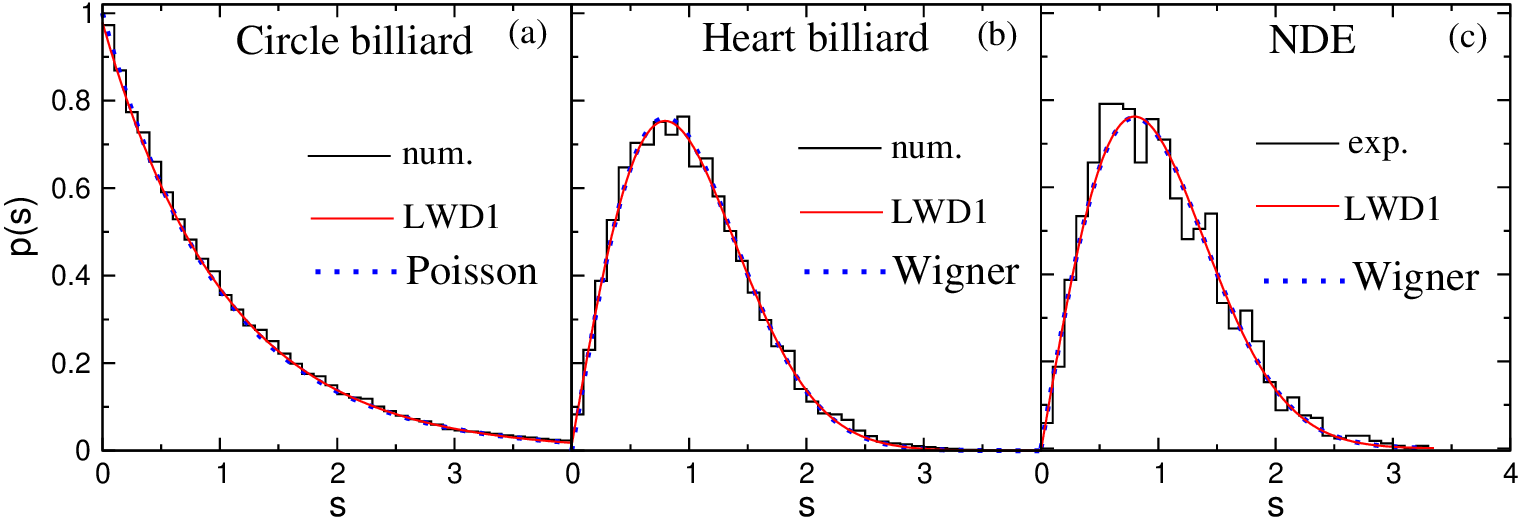}

\vspace{-0.1cm}
\caption{{\small
NNSDs $p(s)$ as functions of a
dimensionless spacing variable $s$ for
(a) Poisson- and (b) Wigner-like numerical calculations and (c)
experimental
Wigner-like results (see text) by staircase lines. LWD1 NNSDs (\ref{pslinWD1})
are shown by solid lines. Dots present the Poisson (a) and Wigner (b,c) curves
(\ref{pois}).}}
\label{fig1}
\end{figure*}

Keeping a link with the analytical properties of the
level repulsion density $g(s)$, it is
convenient to define the probability
$p(s)$ [Eq. (\ref{WDgen})] with
a general smooth
density $g(s)$, that
is a polynomial of not too a large power.
 As shown
in Refs.~\cite{BMprc12,LMRprc17},
this density smoothness is
essentially used in the derivation
of Eq.~(\ref{WDgen}).
For the
simplest statistical analysis in terms of the Poisson- and Wigner-like
distribution contributions, one can use the
expansion of $g(s)$ in series of a few powers of $s$,
\begin{equation}
g(s)\approx a + b s~,
\label{denlin}
\end{equation}
where $a$ and $b$
are fitting parameters. Substituting
this expansion
into the general Wigner-Dyson
formula (\ref{WDgen}),
one obtains explicitly the analytically simple distribution
\begin{equation}
p^{}_{\tt LWD}(s)=(a+b s)~\exp\left(-a s - \frac{b}{2} s^2\right).
\label{pslinPW}
\end{equation}

Taking the limits $a \rightarrow 1$, $b\rightarrow 0$
and $a \rightarrow 0$, $b\rightarrow \pi/2$ in Eq.~(\ref{pslinPW}),
one simply arrives relatively at
the standard Poisson $p^{}_{P}(s)$ and
Wigner  $p^{}_{W}(s)$
distributions (\ref{pois}). In this way,
a linear approximation
(\ref{denlin}) unifies analytically these two limit cases through  a smooth
level repulsion density $g(s)$.
Its parameters
$a$ and $b$ in  Eq.\ (\ref{denlin}) are  associated with
the Poisson and Wigner distribution contributions.

The two-parameter distribution LWD2 (\ref{pslinPW})
can be simplified  approximately by reducing it to
one parameter.  As Eq.~(\ref{pslinPW}) obeys identically
the normalization condition (\ref{normcond1}),
one satisfies only
the approximate normalization condition (\ref{normcond2}).
Thus, one obtains a
relation between the parameters $a$ and $b$
[marked by low index one in $ p^{}_{\tt LWD}(s)$]:
\begin{equation}
\int_0^{\infty}  s\; p^{}_1(s)\;\mbox{d}s \equiv
\sqrt{\frac{\pi}{2b}}~e^{w^2_{}}~\mbox{erfc}(w) = 1\;,
\label{normcond22}
\end{equation}
where
\begin{equation}
w=a/\sqrt{2b}~.
\label{zeta}
\end{equation}
Here, $\mbox{erfc}(w)$ is the standard error
function,
\begin{equation}
\mbox{erfc}(w)=1-\mbox{erf}(w)\equiv
1-\frac{2}{\sqrt{\pi}}\int_0^w \mbox{d}x~\exp(-x^2)~.
\label{erfc}
\end{equation}
Solving Eq.~(\ref{normcond22}) with respect to $b=b(w)$ and using
Eq.~(\ref{zeta}) for $a=a(w)$, one finds
\begin{equation}
p^{}_{1}(s)=[a(w)+b(w) s]~\exp\left[-a(w) s - \frac{b(w)}{2} s^2\right],
\label{pslinWD1}
\end{equation}
where
\begin{equation}
a=\sqrt{\pi}~w~e^{w^2_{}}~\mbox{erfc}(w)~,\qquad
b=\frac{\pi}{2}~e^{2w^2_{}}~~\mbox{erfc}^2_{}(w)~.
\label{abzpar}
\end{equation}
The probability distribution which obeys both normalization conditions
(\ref{normcond1}) and (\ref{normcond2})
is given by Eq.~(\ref{pslinWD1}),
where $a(w)$ and $b(w)$ are functions of only one
parameter $w$ through Eq.~(\ref{abzpar}).
Eq.~(\ref{pslinWD1}) approaches
the Wigner limit for $w\rightarrow 0$
and
the Poisson limit for $w\rightarrow\infty$, respectively,
\begin{equation}
a(w)=\sqrt{\pi}w + O(w^2),~~
b(w)=\frac{\pi}{2} - 2\sqrt{\pi}w + O(w^2)
\label{abzpar0}
\end{equation}
and
\begin{equation}
a(w)=1 -\frac{1}{2w^2}+O\left(\frac{1}{w^3}\right),~~
b(w)=\frac{1}{2w^2}+O\left(\frac{1}{w^3}\right).
\label{abzparinf}
\end{equation}
Thus, the probability density (\ref{pslinWD1}) is a  simple
analytical continuation
 from
the Poisson  $p^{}_{P}(s)$ to Wigner  $p^{}_{W}(s)$ limit distributions
(\ref{pois})
through  a smooth linear
  level-repulsion density $g(s)$ [Eq.~(\ref{denlin})],  and both equations
(\ref{normcond1}) and (\ref{normcond2}) are satisfied.

\section{Discussions of results}
\label{disc}

Fig.~\ref{fig1} shows
the results of testing the LWD1 [Eq.~(\ref{pslinWD1})]
by fitting the NNSDs  with
a good statistics: Numerical
quantum spectra in the circular (a) and heart (b) billiards,  and
for the nuclear data ensemble [NDE, (c)]. The NDE
includes
1726 neutron and proton resonance energies   ~\cite{Haq82}.
The LWD1 (\ref{pslinWD1}) is in good agreement
with both
numerical (a,b) and experimental NDE (c) NNSDs,
as well with the corresponding Poisson (a) and Wigner (b,c) limits, see
Eqs.~(\ref{pois}),
(\ref{abzpar0}), (\ref{abzparinf})
and Table \ref{table1}.
The sampling intervals
for building the NNSDs
(after the unfolding procedure \cite{LMRprc17}) in
Fig.~\ref{fig1}
are given by $\gamma_s=0.1$.
In all other figures, one finds the reliable parameter $\gamma_s=0.2$.
They are taken from the condition of the
stable smoothed NNSD values
without sharp jumps between the neighbor
energies.

\begin{figure*}
\includegraphics[width=0.75\textwidth,clip]{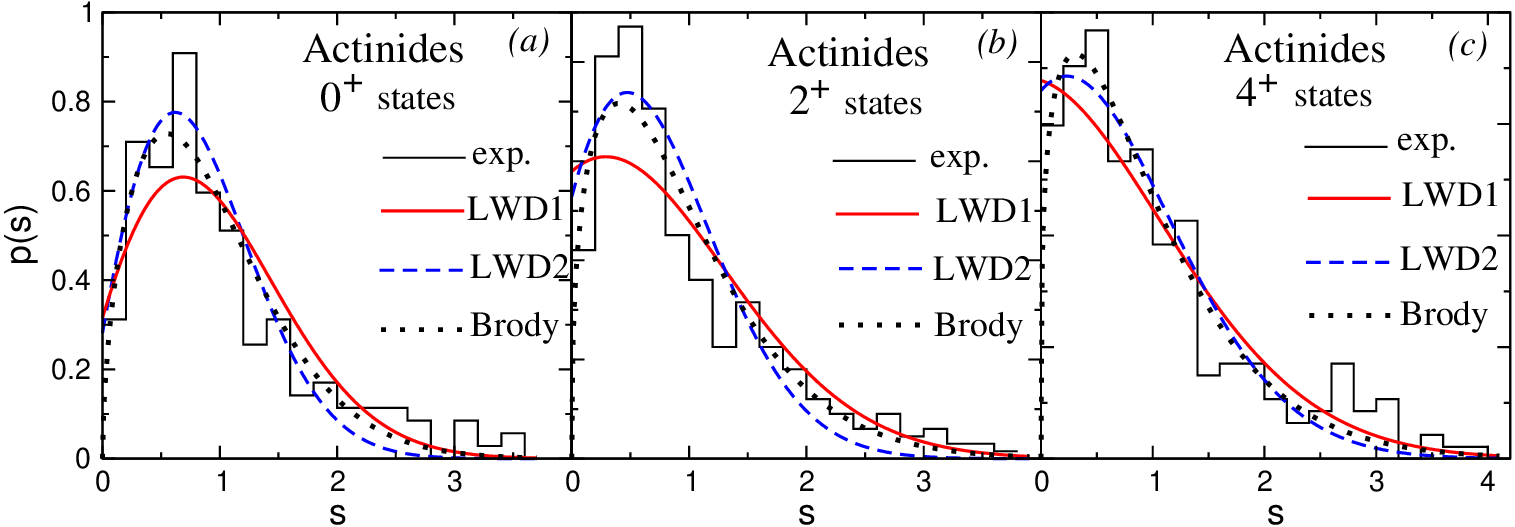}

\vspace{-0.1cm}
\caption{{\small
The same as in Fig.~\ref{fig1} but
for different experimental states in the actinide nuclei;
(a-c):
for 0$^+$, 2$^+$, and 4$^+$,
respectively. The fits
by the LWD1 (\ref{pslinWD1}), LWD2 (\ref{pslinWD2})  and Brody \cite{Bro73}
approach
are respectively
shown by
solid, dashed and dotted lines
(staircase lines in Fig.~\ref{fig2} 
are taken from
Ref.~\cite{LMRprc17}).
}}
\label{fig2}
\end{figure*}

Experimental NNSDs for the collective states
with different angular momenta $I^{\pi}=0^+,2^+,4^+$ are
excited in several actinide nuclei.
They are fitting by the  LWD1 (\ref{pslinWD1})
and  LWD2
approximations (\ref{pslinWD2})
in Fig.~\ref{fig2}, see also
the parameters of these fittings given in
Table \ref{table1}. All
\vspace{-1.0cm}
\begin{table*}[pt]
\begin{center}
\begin{tabular}{|c|c|c|c|c|c|c|c|c|c|}
\hline
~Figure~  &~system~ &~ $a^{}_1$~&
~$b^{}_1$ ~&~ $w$~ &~ $\chi^2_1$ ~& ~ $a^{}_2$~ &~ $b^{}_2$ ~
& $~\langle s\rangle$ ~&~ $\chi^2_2$~\\
\hline
1a & circle & ~0.98~ & ~0.02~ & ~4.79~
&~ 0.99~ &~ 0.98~ & ~0.03~ &~ 0.99~ &~0.9~
\\
~b &  heart & 0.08 & 1.41 & 0.05
& 3.6 & 0.02 & 0.98 & 0.99 &2.1  \\
~c & NDE & 0.07 & 1.44 & 0.04
& 0.99 & 0.00 & 1.03 & 0.99 &5.6\\
\hline
2a & 0$^+$& 0.32  & 0.98 & 0.23
& 11.4& 0.26  & 1.02& 0.85 &9.2 \\
~b & 2$^+$& 0.58 & 0.54 & 0.56
& 11.8& 0.53  & 0.74 & 0.82 & 10.2 \\
~c & 4$^+$& 0.68 & 0.40 & 0.76
& 9.1& 0.66  & 0.42& 0.89 &8.5\\
\hline
3a & 4$^+$~all K& 0.50 & 0.67 & 0.43
& 11.8 & 0.50 & 0.56 & 0.91 & 11.5 \\
~ b & $K=0$ & 0.32 & 0.97 & 0.23
& 11.0 & 0.31 & 0.84 & 0.89 &9.9 \\
~c & $K=2$& 0.07& 1.44
&0.04  & 11.3 & 0.02& 0.86 & 1.06 &10.6 \\
 ~d & $K=4$& 0.14& 1.30
&0.08  & 12.3 & 0.07& 0.77 & 1.08 &11.1 \\
\hline
\end{tabular}

\caption{{\small
Parameters $a_i$, $b_i$ and $w$ of one- and two-parameter
LWDi approximations ($i=1,2$) [Eqs.~(\ref{pslinWD1}) and (\ref{pslinWD2})
for $i=1$ and $2$, respectively]
for the exemplary cases and collective excited states
in several nuclei. Results: Fig.~\ref{fig1}(a) and (b) are the NNSDs given
for the numerical
circle and heart billiard calculations \cite{Gomez11}, (c) presents
NNSDs for many experimental neutron-resonance NDE \cite{Gomez11,Haq82};
Fig.~\ref{fig2}(a-c) is for experiments with actinides;
NNSD parameters of Fig.~\ref{fig3} for all mixed projections of $K$
(a) are
compared with the symmetry breaking
by setting $K=0$ (b), $2$ (c)
and $4$ (d). Averaged $s$  values $\langle s \rangle$
[Eq.~(\ref{normcond2LWD2}] for the LWD2 are shown in the 9th column.
The standard accuracies found by $\chi^2_i$
of least-squares fittings (in percent)
are shown in the 6th and 10th columns.
}}
\label{table1}
\end{center}
\end{table*}
%

\vspace{1.0cm}
\noindent spectra in the same energy interval 0--3 MeV demonstrate
 an intermediate structure between an order and a chaos
 with varying dominance of the Wigner to the
Poisson contribution for increasing the angular momentum from 0$^+$ to 4$^+$.
With increasing angular momentum, one can see a shift of the NNSD to the
Poisson limit.
As shown in Ref.~\cite{LMRprc17}, e.g., spectra 0$^+$
in the energy interval 0--3 MeV
are intermediate
between an  order and a  little more pronounced
chaos structure, while
ordered nature is  dominant for the experimental
spectra in the extended energy interval about 0−-4 MeV.

%
\begin{figure*}
\includegraphics[width=0.6\textwidth,clip]{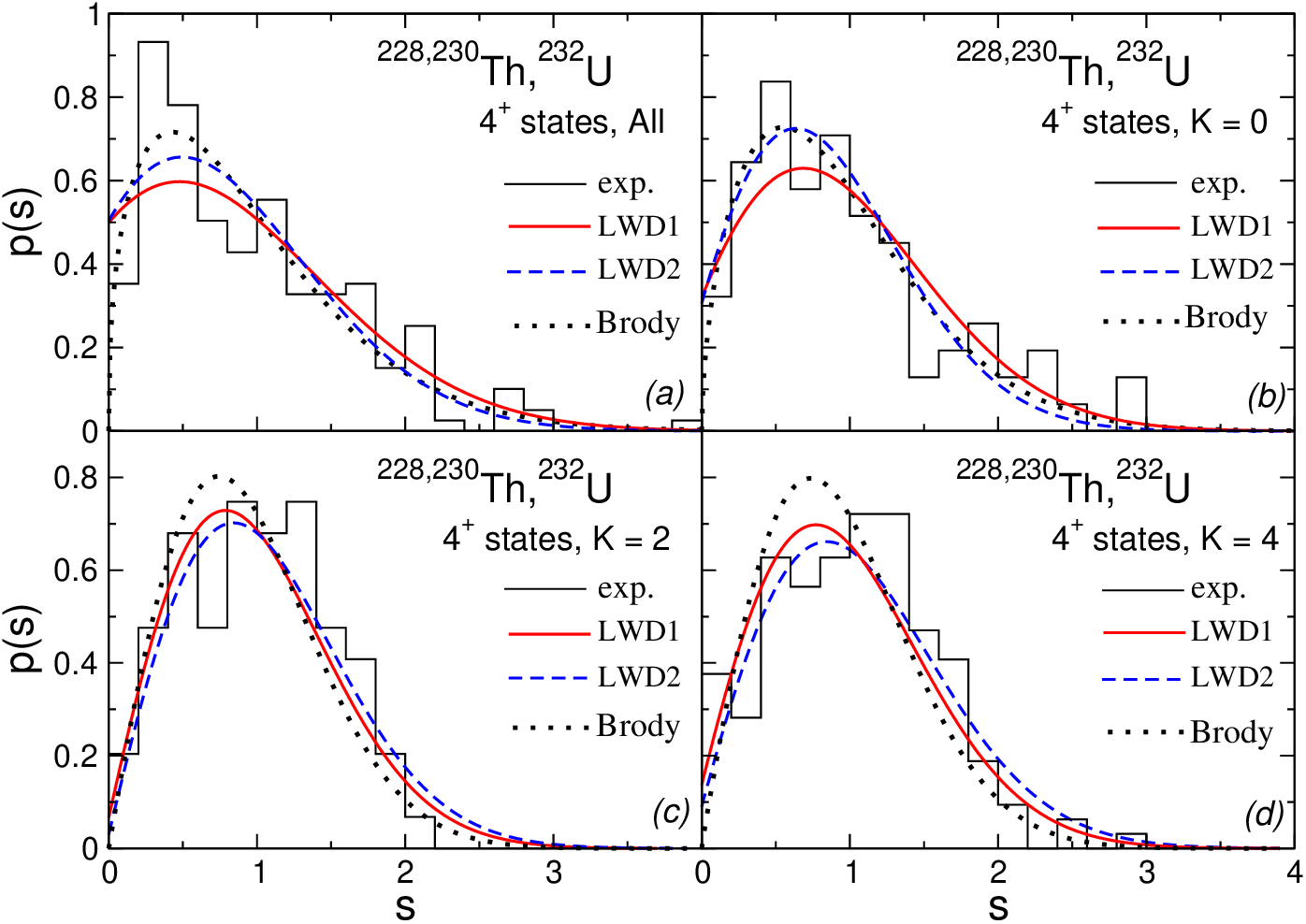}

\vspace{-0.1cm}
\caption{{\small
NNSDs for full spectrum (a) and symmetry breaking by fixed $K=0$ (b),
$2$ (c) and $4$ (d) projections of the
angular momentum $4^+$ for the
 actinides which are included
in Fig.~\ref{fig2}. Solid, dashed and dotted lines are fits by the
LWD1 (\ref{pslinWD1}), LWD2 (\ref{pslinWD2})  and Brody \cite{Bro73} NNSDs,
respectively.
}}
\label{fig3}
\end{figure*}
Taking actinides as example, see Ref.~\cite{LMRprc17},
one shows
a good agreement between
experimental \cite{Lev09,Lev13,Lev15}
and theoretical \cite{Sol92} results that confirms the
collective nature of the
desired states and, finally,
completeness of the level sequences. At the same time, the
theoretical
distribution for the extended energy
interval 0-4 MeV
is shifted to the Poisson law
as compared to the  experimental and theoretical
distributions in the interval 0-3 MeV.

Fig.~\ref{fig3} shows the typical example of
symmetry breaking phenomena \cite{Gomez11}.
Experimental NNSDs are obtained from the analysis of sequences
for the 4$^+$ states performed for three actinides $^{228,230}$Th and $^{232}$U
\cite{Lev09,Lev13,Lev15}. The rotational bands were
identified in this analysis.
The identification of states
that is associated with rotational bands was
performed on the following conditions.
i) The angular distribution for
the state with a given spin as a band member
candidate is
assigned by the DWBA calculations.
(This state is necessarily included into the band.)
ii) The transfer cross section in the (p,t) reaction to
states in the potential band has to be decreased with
increasing spin.
iii) Energies of the states in
this band can be
evaluated approximately by the expression for a rotational band
$E = E_0 + {\mathcal A}I(I + 1)$ with a small and smooth variation
of the inertial parameter $\cal{A}$.
In such a way new sequences with the angular
momentum $4^+$ were formed, separately for the $K = 0,~ 2$ and 4 states.
The latter can be considered as pure sequences.
As seen from
Fig.~\ref{fig3},
if we fix the projection of the angular momentum
$K$, the NNSD is changed toward the chaos (Wigner distribution) for
each of the cases $K=0$
 [Fig.~\ref{fig3}(b)] versus
$2$ (c) and $4$ (d) (see also Table \ref{table1}).
This effect observed in both LWD1 and LWD2 approximations
 is more enhanced for the $K=2$ (c) [or $4$ (d)] case than for $K=0$ (b).
Thus, we found a similar effect of the symmetry breaking
like in the single-particle spectra \cite{BMprc12}, that was explained
by decreasing the number of single-valued integrals of motion.
This analysis confirms also a proposed explanation of NNSD
shifts to the
Poisson limit with increasing the angular momentum
$0^+$, $2^+$ and $4^+$ (Fig.~\ref{fig2})
by mixing
the sequences with different symmetries (different $K$).
An extended version of the
proper discussions of these phenomena will be presented in
the forthcoming work.

As seen from Figs.~\ref{fig2}
and \ref{fig3} and Table \ref{table1},
results of the
fitting of experimental NNSDs
\cite{Lev94,Lev09,Lev13,Lev15,Spi13,Les02,Buc06,Mey06,Bet09,Spi17}
by the LWD1 (\ref{pslinWD1}),
LWD2 (\ref{pslinWD2})  and Brody approach \cite{Bro73} are basically close,
though some differences are visible.
Their main features,
- the position of maxima and the Poisson $a$ and
Wigner $b$ distribution contributions, - are
approximated within good accuracy
of calculations. In
the LWD1 approximation we related
 these parameters by
satisfying the normalization condition (\ref{normcond2}) which is
idealized
as compared to Eq. (\ref{normcond2LWD2}) with respect to
the upper integration limit
in the LWD2 approach (Sec.~II and Appendix).
As a result, the LWD1 has one parameter for fitting 
  as the Brody NNSD.
In these LWD1 derivations we assumed
a fast convergence of the
normalization integral in Eq.~(\ref{normcond2LWD2}) as function
of  a maximal spacing
value $s^{}_{\rm max}$.
On the other hand, in the LWD2 case [Eq.~(\ref{pslinWD2})]
we keep $a$ and $b$ independent in the fitting procedure and check, then,
the accuracy of Eq.~(\ref{normcond2LWD2}) for $\langle s \rangle$
(Table \ref{table1}).
The upper integration limit $s_{\rm max}$
must be larger than all of energy spacings
in a given experimental spectrum,
and this has to be checked too. The LWD2 approximation (\ref{pslinWD2})
 and Brody formula (besides of small values of $s$)
look visually better fitted with
the improved accuracy (see
Figs.~\ref{fig2}
and \ref{fig3} and Table ~\ref{table1}),
especially
near maxima of the
experimental data. 
The  LWD1 is better fitted on the right of the distribution maximum
in a wider spacing interval.
This provides
explicitly the
normalization condition (\ref{normcond2}).
A simpler one-parameter fitting has obvious analytical
advantages. In particular, the LWD1 is preferable for calculations
in Fig.~\ref{fig2}(b)
where the LWD2
average
$\langle s \rangle$  (\ref{normcond2LWD2}) differs
notably from one.
(Table \ref{table1}).
Thus, sometimes, the  LWD1 and LWD2
NNSDs can be
helpful as those giving a complementary
information on statistical properties of quantum spectra.

\section{Conclusions}
\label{concl}

We derived the
simple one-parameter NNSD approximation to the
Wigner-Dyson probability distribution.
 Several exemplary problems
were demonstrated: standard
circular (Poisson) and cordial (Wigner) billiards,
and famous experimental neutron-resonance
states in many
nuclei (Fig.~\ref{fig1}). Using this approximation
we provide
statistical analysis of
the nuclear collective excitations with several
spins (Fig.~\ref{fig2}):
0$^+$, 2$^+$, and
4$^+$ in a number of actinides
to show the
good agreement with the one-parameter LWD1,
as well as with the two-parameter LWD2 (Table \ref{table1}).
For the
linear approximation to
level repulsion densities, one obtains a clear
information on the
quantitative measure of the Poisson order and Wigner chaos contributions
in the experimental spectra,
separately, in contrast to the heuristic
Brody approach.  However, one finds in
our calculations
that the Brody formula
agrees largely well with
the LWD probability-distribution results (again, 
apart from small values of $s$).
The precision of fitting for the experimental data by
the two-parameter LWD2 is
improved but the full analytical
 one-parameter LWD1 approximation has an obvious advantage.
 Simplifying analytically the
normalization
condition for the spacing average
we do not need to check
its precision.

We confirm the intermediate structure
between the Poisson and Wigner statistical peculiarities of the
experimental spectra for nuclear collective states
by evaluating their separate contributions
(Fig.~\ref{fig2}).
Also, one finds
that the Wigner contribution dominates in the NNSD
for 0$^+$ states and the Poisson contribution is
enhanced with increasing the angular momentum.
All considered nuclear spectra
are collective and complete for a given angular momentum $I$ and parity
$\pi$.
In accordance with Ref.~\cite{Gomez11}, for
collective states of a
wider energy interval in deformed nuclei,
the statistical distributions
are closer to
the Poisson distribution, and in other cases the situation is intermediate
(see also Ref.~\cite{Shrin91}),
in contrast to the single-particle states
 \cite{Dietz17}.
It has been shown that the symmetry breaking due to the fixing of the
projection $K$ of angular momentum $I$ enhances the chaos by
a shift of the NNSD toward the Wigner distribution.
This property is
common for the collective
 and single-particle states.
In perspective, it will be worthwhile to study more systematically
the influence of
symmetry breaking
phenomena on these distributions of the collective states
in deformed nuclei.

\vspace{0.1cm}
\centerline{{\bf Acknowledgments}}

\vspace{0.1cm}
We are grateful
to K.~Arita, S.~Aberg, J.P.~Blocki,
S.~Mizutori, K.~Matsuyanagi, V.A.~Plujko, and P.~Ring,
for many helpful discussions.
One of us (A.G.M.) is also very grateful for the kind hospitality during his
working visits of Physical Department of the Nagoya Institute of Technology,
also the Japanese Society of Promotion of Sciences for financial support,
Grant No. S-14130.

\vspace{0.2cm}
\setcounter{equation}{0}
\renewcommand{\theequation}{A\arabic{equation}}

\begin{center}
\textbf{Appendix: The LWD2 NNSD approach}
\label{appA}
\end{center}

\vspace{0.2cm}

For the comparison, let us present also the LWD2 NNSD approximation
\cite{LMRprc17},
\begin{equation}
p^{}_2(s)=\frac{a+b s}{\aleph}\;
\exp\left(-\frac{b}{2} s^2 - a s\right)\;,
\label{pslinWD2}
\end{equation}
where
\begin{equation}
 \aleph\equiv \int_0^{s^{}_{\rm max}} \mbox{d}s \;
\exp\left(-\frac{b}{2} s^2 - a s\right)=a\aleph_0+b\aleph_1\;,
\label{N}
\end{equation}
\begin{eqnarray}
\aleph_0
&=&
\sqrt{\frac{\pi}{2b}}\;\exp\left(\frac{a^2}{2 b^2}\right)\;
\left[\mbox{erf}\left(\frac{a+b s^{}_{\rm max}}{\sqrt{2b}}\right)-
\mbox{erf}\left(\frac{a}{\sqrt{2b}}\right)\right],
\nonumber\\
\aleph_1
&=&
\frac{1}{b}\left[1-\exp\left(-\frac{b}{2} s_{\rm max}^2 -
a s^{}_{\rm max}\right)- a\;\aleph_0\right]\;,
\label{pslin}
\end{eqnarray}
 with independent parameters  $a$ and $b$.
As referred to quantum spectra given in a
finite integration limit $s^{}_{\rm max}$,
the LWD2 distribution $p^{}_2(s)$ obeys the
following normalization conditions:
\begin{equation}
\int_0^{s^{}_{\rm max}} p^{}_2(s)\;\mbox{d}s=1
\label{normcond1LWD2}
\end{equation}
and
\begin{equation}
\langle s \rangle \equiv \int_0^{s^{}_{\rm max}}  s\; p^{}_2(s)\;\mbox{d}s = 1\;.
\label{normcond2LWD2}
\end{equation}

\end{document}